\documentclass{PoS}

\usepackage{graphicx, slashed, listings, fancyhdr, amsmath, amsthm, amssymb, color, subfigure}
\usepackage{amstext}
\usepackage{nccmath}
\usepackage{amsbsy}
\usepackage{amsfonts}
\usepackage{mathrsfs}
\usepackage[dvips]{lscape}

\usepackage{setspace}
\usepackage{cite}
\usepackage{epsfig}
\usepackage[normalem]{ulem}
\usepackage{color}
\usepackage{multirow}
\usepackage{array,ragged2e}
\usepackage{calc}
\usepackage[utf8]{inputenc}
\usepackage[shortlabels]{enumitem} 
\usepackage{xcolor}
\usepackage{dcolumn}
\usepackage{xspace}

\def\Fig#1{Fig.~\ref{#1}}

\newcommand{\twofigs}{0.49\linewidth}

\title{Center Phase Transition from\\ Fundamentally Charged Matter Propagators}

\ShortTitle{Center Phase Transition from Matter Propagators}

\author{\speaker{Mario Mitter}\\
        Universit\"{a}t Heidelberg, Institut f\"{u}r Theoretische Physik, Philosophenweg 16, D-69120 Heidelberg, Germany\\
        E-mail: \email{mario.mitter@thphys.uni-heidelberg.de}}
        
\author{Markus Hopfer\\
       Institut f\"{u}r Physik, Karl-Franzens-Universit\"{a}t Graz, Universit\"{a}tsplatz 5, 8010 Graz, Austria\\
       E-mail: \email{markus.hopfer@uni-graz.at}}
       
\author{Bernd-Jochen Schaefer\\
       Institut f\"{u}r Physik, Karl-Franzens-Universit\"{a}t Graz, Universit\"{a}tsplatz 5, 8010 Graz, Austria\\
       E-mail: \email{bernd-jochen.schaefer@uni-graz.at}}
       
\author{Reinhard Alkofer\\
       Institut f\"{u}r Physik, Karl-Franzens-Universit\"{a}t Graz, Universit\"{a}tsplatz 5, 8010 Graz, Austria\\
       E-mail: \email{reinhard.alkofer@uni-graz.at}}

     \abstract{The center phase transition at non-vanishing
       temperatures is investigated in Landau gauge Quantum
       Chromodynamics (QCD) and scalar QCD. For each theory novel
       order parameters for the transition are introduced. The
       matter-gluon vertex which occurs in the Dyson-Schwinger
       equations of the propagators has to be modeled in
       contemporary studies.  It is found that the nature of the phase
       transition depends strongly on the detailed structure of this
       vertex.
       Our investigation motivates a precise determination of the
       matter-gluon vertex at non-vanishing temperatures.}

\FullConference{Xth Quark Confinement and the Hadron Spectrum,\\
		October 8-12, 2012\\
		TUM Campus Garching, Munich, Germany}

\begin{document}

\section{Introduction}

Investigations of the phase structure of strongly-interacting matter
have received a considerable amount of attention in the last years,
both, theoretically as well as experimentally. Among the most
prominent features of the QCD phase diagram is the crossover from a
confined phase with spontaneously broken chiral symmetry to a chirally
symmetric and deconfined phase.  A confined phase can be linked to a
ground state that respects center symmetry,\footnote{In a strict sense
  center symmetry is realized only in the limit of infinitely heavy
  quarks while in real QCD the symmetry is always explicitly broken,
  see e.g.~\cite{Polyakov:1978vu, Karsch:2003jg}.}
and the associated order parameter is the Polyakov loop. It vanishes
in the center symmetric phase and becomes finite as soon as the center
symmetry is broken \cite{Polyakov:1978vu}. At vanishing quark chemical
potential both transitions occur roughly at the same temperature which
led to the idea and introduction of new dual observables in lattice
QCD \cite{Gattringer:2006ci, Bruckmann:2006kx,Bilgici:2009tx,
  Synatschke:2007bz}.  In particular, the dual chiral condensate
pioneered in \cite{Gattringer:2006ci} is constructed from the chiral
condensate, the order parameter of chiral symmetry breaking. More
recently, dual observable have also become accessible within
functional methods \cite{Fischer:2009wc, Fischer:2009gk, Braun:2009gm}
and have been successfully applied to investigate the center
transition.  In the present work novel order parameters for the center
symmetry and its breaking are introduced and analyzed in QCD as well
as fundamentally charged scalar QCD. The order parameters are
determined by the corresponding matter propagators without any
additional renormalization.

\section{(Scalar) Quantum Chromodynamics}

We address the deconfinement transition in ordinary QCD as well as in
scalar QCD, where the quarks are replaced by fundamentally charged
scalars (see e.g. \cite{Fister:2010yw}) both in Landau gauge.

\begin{figure}[t!]

{\centering

\begin{tabular}{ m{2.0cm} m{0.5cm} m{2.0cm} m{0.5cm} m{2.0cm} m{0.5cm}}
\centering
    \includegraphics[width=2cm]{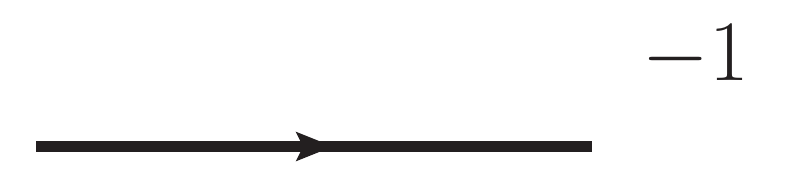}
&
$=$
&
\centering
    \includegraphics[width=2cm]{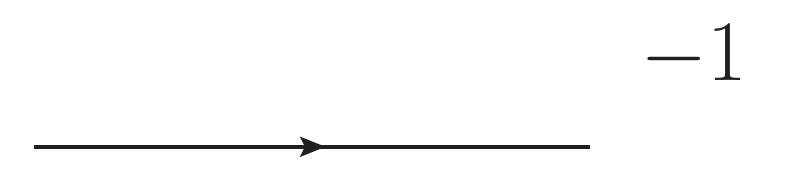}
&
$-$
&
\centering
    \includegraphics[width=2cm]{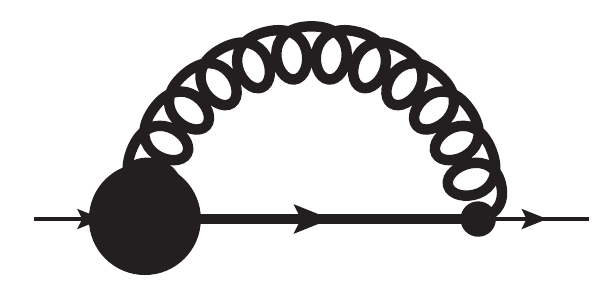}
&

\end{tabular}

}

\caption{DSE for the quark propagator.}
\label{fig:prop_dse}
\end{figure}

The matter propagators are calculated by means of the corresponding
Dyson-Schwinger equation (DSE) \cite{Alkofer:2000wg}.  As an example,
the DSE for the quark propagator is shown diagrammatically in
\Fig{fig:prop_dse}, where thin lines and dots represent bare
propagators and one-particle irreducible vertices while thick lines
and dots denote the corresponding dressed quantities.  Accordingly,
the DSE for the scalar propagator, which is shown in
\Fig{fig:dse_scalprop}, is more involved and contains more diagrams
due to the presence of additional bare vertices such as the scalar
self-interaction and the quartic scalar-gluon vertices. In one-loop
approximation only the momentum-independent tadpole diagrams are left
in addition to the gluon exchange diagram. The tadpoles, however,
can be treated by adjusting the renormalization constants
appropriately. Hence, in a one-loop approximation, the DSE for the
scalar propagator is of the same structure as the one for the quark
propagator, cf.~\Fig{fig:prop_dse}.

Explicitly, at finite temperature $T$ the DSE for the quark propagator
$S(p)$ reads
\begin{equation}\label{eq:quarkDSE}
 S^{-1}(p) = Z_2 S_0^{-1}(p) - Z_{1F}\,C_F\, g^2 T \sum_{\omega_k(\theta)} 
 \int\frac{d^3k}{(2\pi)^3}\, \gamma^\mu S(k) \Gamma^\nu(k,p;q) D^{\mu\nu}(q) 
\end{equation}
and correspondingly for the scalar propagator $D_S(p)$
\begin{equation}\label{eq:scalarDSE}
 D_S^{-1}(p) = \hat Z_2(p^2+\hat Z_m m_0^2) - \hat Z_{1F}C_Fg^2 T \sum_{\omega_k(\theta)}
 \hspace{-0.05cm}\int\hspace{-0.15cm}\frac{d^3k}{(2\pi)^3}(p+k)^\mu
 D_S(k) \Gamma_S^\nu(k,p;q) D^{\mu\nu}(q) \ .
\end{equation}
For the four-momenta we use $k=(\vec{k},\omega_k(\theta))$ and the
gluon momentum is constrained by momentum conservation to $q=p-k$.
The wave function renormalization of the quark (scalar) fields are
denoted by $Z_2$ ($\hat Z_2$), the renormalization constants of the
quark-gluon (scalar-gluon) vertex by $Z_{1F}$ ($\hat Z_{1F}$) and
$Z_m$ labels the scalar mass renormalization constant.  The quadratic
Casimir invariant in the fundamental representation of the gauge group
$SU(3)$ is $C_F=4/3$ and the coupling constant at the renormalization
scale is given by $g$.  

In general, $\exp(i\theta)$-valued boundary conditions in the fourth
spacetime direction are realized by introducing generalized Matsubara
frequencies $\omega_k(\theta)=(2\pi n_k + \theta)T$, where the sums
over $\omega_k(\theta)$ in the DSEs run over the corresponding
discrete values $n_k\in\mathbb{Z}$.  The usual (anti-)periodic
boundary conditions for (fermions) bosons are obtained by setting
($\theta=\pi$) $\theta=0$, respectively.

\begin{fleqn}

\begin{figure}[t!]

{\centering

\begin{tabular}{ m{2.0cm} m{0.35cm} m{2.0cm} m{0.35cm} m{2.0cm} m{0.35cm} m{2.0cm} m{0.35cm} m{2.0cm} m{0.35cm}}

\centering
    \includegraphics[width=1.5cm]{s2_inv}
&
$=$
&
\centering
    \includegraphics[width=1.5cm]{s2_binv}
&
$-$
&
\centering
    \includegraphics[width=1.5cm]{s2_A1s2}
&
$-\ \frac{1}{2}$
&
\centering
    \includegraphics[width=1.5cm]{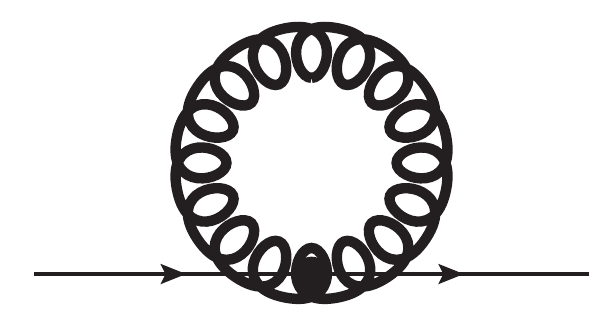}
&
$-$
&
\centering
    \includegraphics[width=1.5cm]{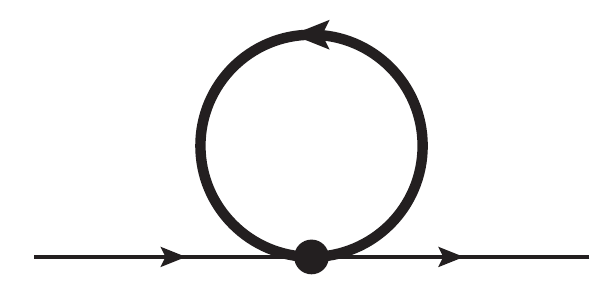}
&

\\

&
$-\ \frac{1}{2}$
&
\centering
    \includegraphics[width=1.5cm]{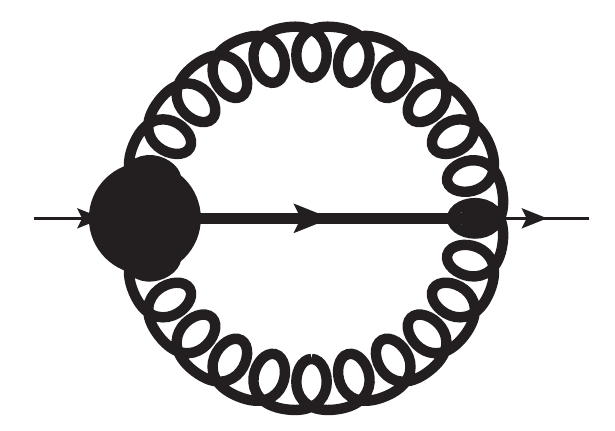}
&
$-\ \frac{1}{2}$
&
\centering
    \includegraphics[width=1.5cm]{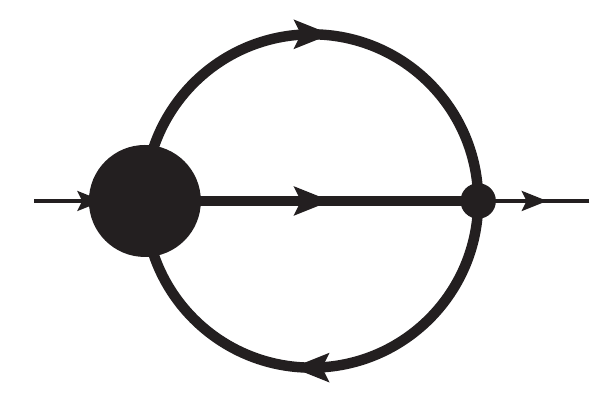}
&
$-$
&
\centering
    \includegraphics[width=1.5cm]{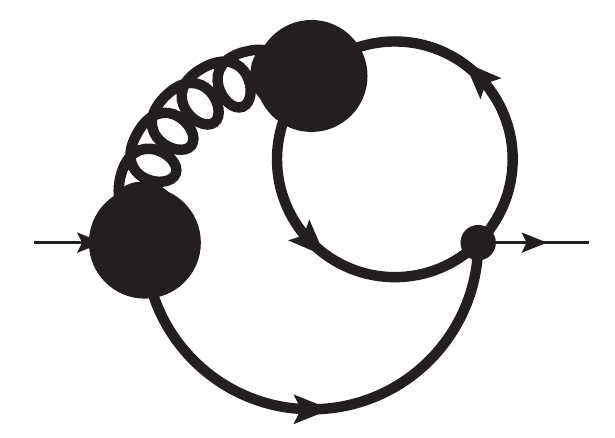}
&
$-\ \frac{1}{2}$
&
\centering
    \includegraphics[width=1.5cm]{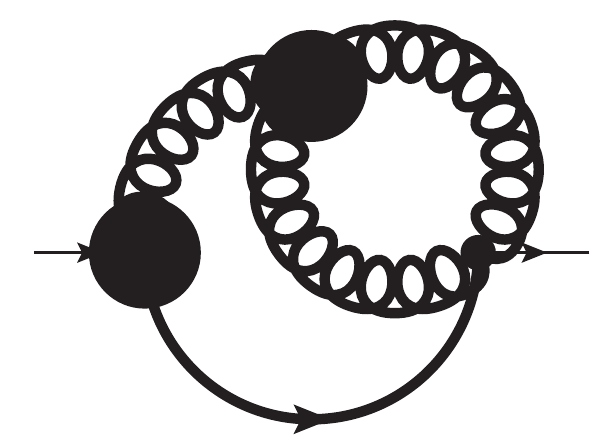}
&

\\

&
$-$
&
\centering
    \includegraphics[width=1.5cm]{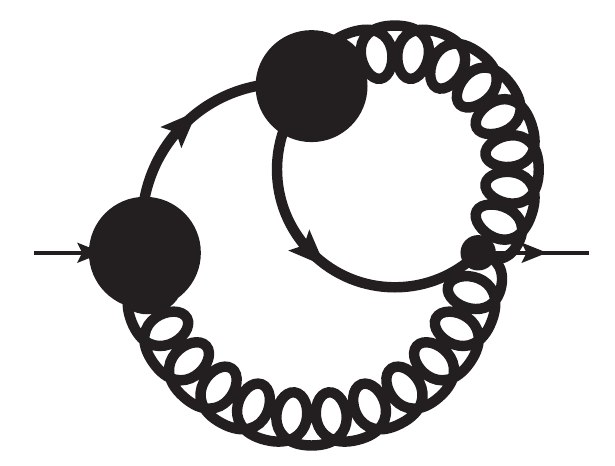}
&

&

&

&

&

&

\end{tabular}

}

\caption{DSE for the fundamentally charged scalar propagator.}
\label{fig:dse_scalprop}
\end{figure}

\end{fleqn}

Both DSEs depend on the gluon propagator $D^{\mu\nu}$ as well as on
the corresponding matter-gluon vertices $\Gamma^\nu$ and
$\Gamma_S^\nu$. Solutions of the DSE for the Landau gauge gluon propagator 
at non-vanishing temperatures have been obtained in \cite{Maas:2004se}. 
In addition, data are available from
(quenched) lattice simulations at finite temperatures and have already
been successfully implemented in functional equations for the quark
propagator \cite{Fischer:2009gk,Fischer:2010fx}\footnote{Recently,
  also unquenched lattice data for the Landau gauge gluon propagator
  are available \cite{Aouane:2012bk}, which will be used in future
  studies of the system.}.
In this work we will apply the fit functions for the gluon propagator
proposed in \cite{Fischer:2010fx} and hence we omit explicit
expressions.  

For the matter-gluon vertices the situation is less
satisfactory since the temperature behavior of these vertices is not
known generally. Some modeling has to be employed such as in
\cite{Fischer:2009gk} where the following expression
\begin{equation}
 \begin{split}
   \Gamma^\nu(k,p;q) = \tilde Z_3 &
   \biggl(\delta^{4\nu}\gamma^4\frac{C(k)+C(p)}{2} +
   \delta^{j\nu}\gamma^j\frac{A(k)+A(p)}{2} \biggr) \\
   & \times\Biggl\{\frac{d_1}{d_2+q^2} +
   \frac{q^2}{q^2+\Lambda^2}\biggl( \frac{\beta_0\alpha(\mu)
     \ln\left[q^2/\Lambda^2+1\right]} {4\pi}\biggr)^{2\delta}\Biggr\}\
 \end{split}
\label{eq:quarkgluonvertex}
\end{equation}
for the quark-gluon vertex can be found and will be used also in this
work. The Ansatz is motivated by Slavnov-Taylor identities of the
Abelian gauge theory (see e.g. \cite{Ball:1980ay}) and by the running
of the non-perturbative coupling of the Yang-Mills theory. The purely
phenomenological parameters $d_1$ and $d_2$ are specified in
\cite{Fischer:2010fx}, whereas the anomalous dimension
$2\delta=-18/44$ and $\beta_0 = 11N_c/3$ ensure a correct
perturbative running coupling in the UV regime for $SU(N_c)$ gauge
theory. The renormalization scale of the Yang-Mills sector has been
fixed by $\alpha(\mu) = 0.3$ and $\Lambda=1.4$ GeV. The factor $\tilde
Z_3$ allows to apply the Slavnov-Taylor identity $Z_{1F}=Z_2/\tilde
Z_3$ in Landau gauge \cite{Taylor:1971ff} which yields finally only a
dependence on the quark wave function renormalization $Z_2$. The gluon
momentum is denoted by $q$ and $k$ and $p$ are the in- and outgoing
quark momenta, respectively.

In a similar context we employ
\begin{equation}\label{eq:scalargluonvertex}
 \begin{split}
 \Gamma_S^\nu(k,p;q) & = \tilde Z_3 \frac{D_S^{-1}(p^2)-D_S^{-1}(k^2)}{p^2-k^2} (p+k)^\nu\\ 
 & \times d_1\Biggl\{\frac{\Lambda^2}{\Lambda^2+q^2} + \frac{q^2}{q^2+\Lambda^2}\biggl(\frac{\beta_0\alpha(\mu)\ln\left[q^2/\Lambda^2+1\right]}{4\pi}\biggr)^{2\delta}\Biggr\}
 \end{split}
\end{equation}
for the scalar-gluon vertex, where one additional parameter $d_1=0.53$
has been introduced and all remaining parameters are the same as in
the quark-gluon vertex.  In contrast to the quark-gluon vertex, the
scalar-gluon vertex is dressed only with the vacuum propagators
$D_S^{-1}(p^2)$.  For the numerical solution of the corresponding DSEs
we rewrite the propagators $S^{-1}(p) = i\gamma_4\omega_p(\theta) C(p)
+ i\slashed{\vec p} A(p) + B(p)$ and $D_S(p)=Z_S(\vec{p}^{\,
  2},\omega_p(\theta))/(\vec{p}^{\, 2}+\omega_p(\theta)^2)$ in terms
of dressing functions in a standard way.  Details on the numerical
implementation as well as on the renormalization scheme will be
published elsewhere \cite{our_paper}, cf. also \cite{Huber:2011xc,
  Hopfer:2012qr}.

\begin{figure*}[t!]
  \centering 
  \subfigure{\includegraphics[width=\twofigs]{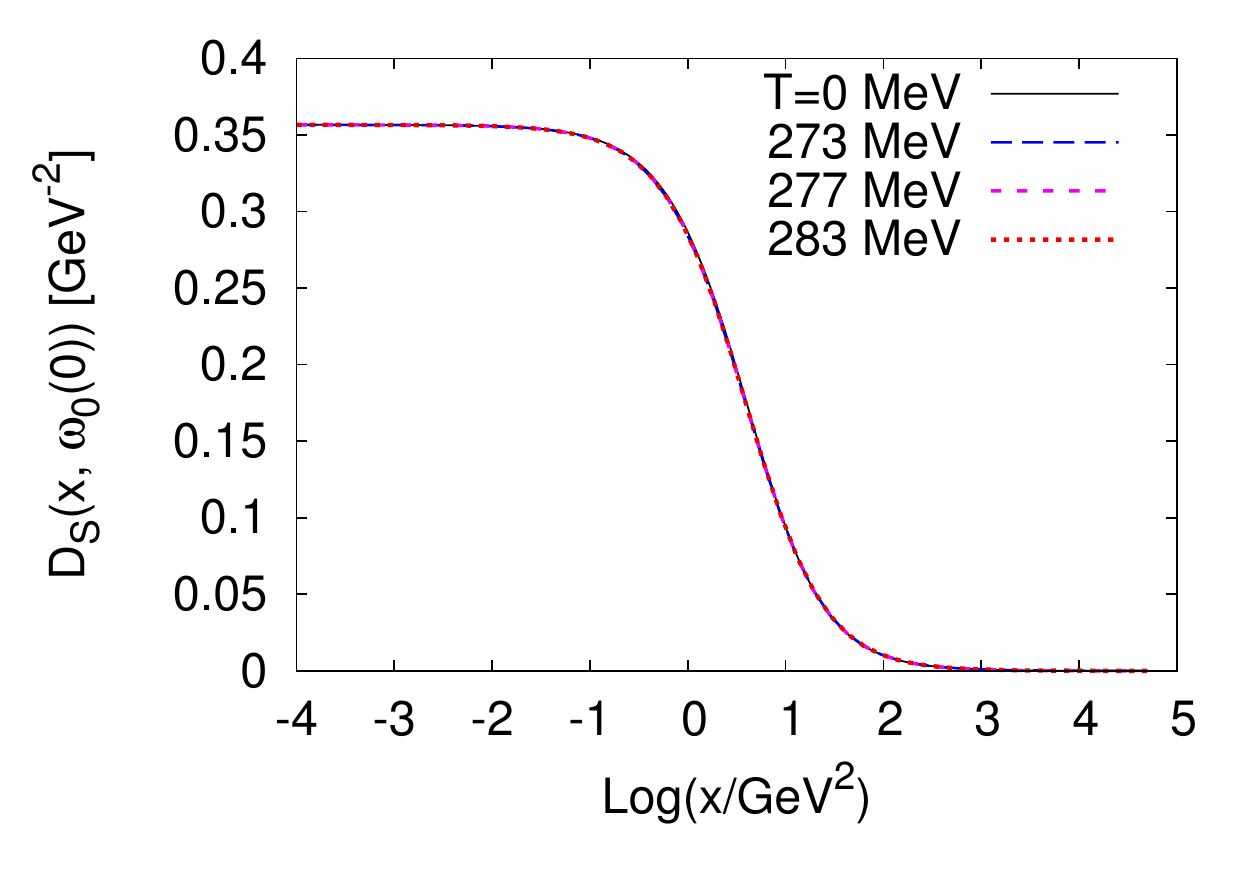}}
  \subfigure{\includegraphics[width=\twofigs]{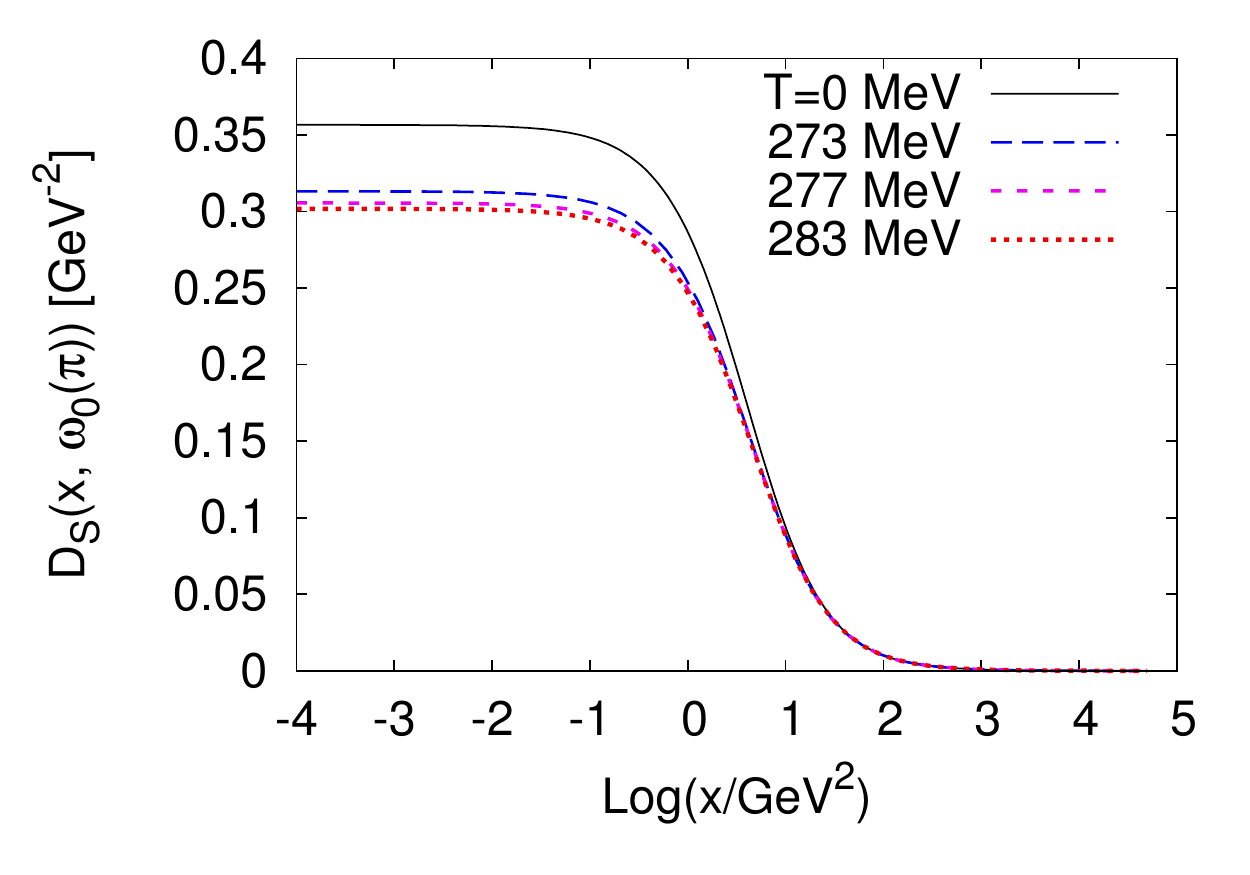}}
  \caption{\label{fig:dse_scalprop_1_T} Solution of the truncated DSE Eq.~\protect\eqref{eq:scalarDSE}
    for the scalar propagator as function of 
    $\vec{p}^{\, 2}=x$ at the lowest Matsubara frequency for periodic $\theta = 0$ (left panel) and antiperiodic $\theta = \pi$ (right panel) boundary conditions at
    different temperatures with
    renormalization scale $\mu = 4$ GeV and mass $m = 1.5$ GeV.}

\end{figure*}

Numerical results for the scalar propagator around temperatures of the
center transition in the quenched theory with $T_c \approx 277$ MeV
are shown in \Fig{fig:dse_scalprop_1_T} for periodic (left panel) as
well as antiperiodic boundary conditions (right panel). The mass of
the scalars has been fixed to $m=1.5$ GeV which results in a quite
inert behavior around $T_c$.
These results demonstrate that there are no direct modifications of
the scalar propagator in the vicinity of the transition. This
motivates the construction of more sensitive order parameters for the
center phase transition.

\section{Center Phase Transition and Dual Order Parameters}

Order parameters for the center phase transition can be composed with
functional methods via dual observables like e.g. the dual chiral
condensate \cite{Fischer:2009wc, Fischer:2009gk} or the dual density
\cite{Braun:2009gm}. As has been discussed in the latter reference,
dual quantities can be evaluated in two different ways.  In general,
dual order parameters are constructed from some boundary-condition
dependent operator $\hat O_{\theta}$, where $\theta$ denotes the phase
of the $U(1)$-valued boundary conditions.  Originally, such operators
were introduced in lattice calculations \cite{Gattringer:2006ci,
  Bruckmann:2006kx, Synatschke:2007bz, Bilgici:2009tx} and evaluated
in QCD with the original boundary conditions, i.e., with
(anti-)periodic boundary conditions for (fermions) bosons. 

On the other hand, these operators can also be evaluated with
functional methods in various theories with general boundary
conditions, referred to as QCD$_\theta$ in \cite{Braun:2009gm}.  
However, dual observables evaluated in this way can serve as order
parameters only if the deconfinement transition temperature at
physical boundary conditions is a lower bound for the transition
temperatures in the different theories QCD$_\theta$
\cite{Braun:2009gm}. 

In previous studies of the center phase transition the dual chiral
condensate has been calculated with functional methods based on the
quark propagator \cite{Fischer:2009wc,
  Fischer:2009gk,Fischer:2010fx}. It is given by an expansion in
complex Fourier modes 
\begin{equation}\label{eq:dualquarkcondensate}
 \Sigma_n = \int_0^{2\pi}\frac{d\theta}{2\pi}
 e^{-in\theta}\langle\bar\psi\psi\rangle_\theta\ 
\end{equation}
with a $\theta$-dependent quark condensate
\begin{equation}
 \langle\bar\psi\psi\rangle_\theta = Z_2 N_c T \sum_{\omega_p(\theta)}\int\frac{d^3p}{(2\pi)^3}\,
 tr_D\,S(\vec p,\omega_p(\theta))\ .
\label{eq:quarkcondensate}
\end{equation}
For arbitrary $n$ that is not a multiple of $N_c$, $\Sigma_n$ can then
serve as an order parameter for center symmetry.  Usually, the dual
chiral condensate $\Sigma_1$ is used, also called dressed Polyakov
loop. In a lattice formulation it contains contributions from all
time-like loops around the torus with winding number $n=1$
\cite{Gattringer:2006ci, Bruckmann:2006kx, Synatschke:2007bz} and
transforms similar to the ordinary Polyakov loop
\cite{Polyakov:1978vu} under center transformations.

Hence, the calculation of the dual chiral condensate requires the
chiral condensate with general boundary conditions, which has to be
regularized for non-vanishing quark masses. As an already finite
alternative we propose
\begin{equation}\label{eq:orderparameter_quark}
  \Sigma_Q =
  \int_0^{2\pi}\frac{d\theta}{2\pi}e^{-i\theta}\;\Sigma_{Q,\theta} \;,\quad
  \Sigma_{Q,\theta} =
  T\sum_{\omega_p(\theta)}\;\left[\frac{1}{4}\ tr_D\;S(\vec
    0,\omega_p(\theta))\right]^2 
\end{equation}
as an order parameter for the center phase transition.
$\Sigma_{Q,\theta}$ is finite, because the sum scales like
$1/\omega_p^4$ for large Matsubara modes. Similarly, for scalar QCD we
propose the order parameter
\begin{equation}
 \Sigma_S = \int_0^{2\pi}\frac{d\theta}{2\pi}e^{-i\theta}\;\Sigma_{S,\theta} \;,
 \quad \quad \Sigma_{S,\theta} = T\sum_{\omega_p(\theta)}D_S^2(\vec
 0,\omega_p(\theta))\ .
 \label{eq:orderparameter_scalar}
\end{equation}
More details on these order parameters will be presented in an upcoming
publication \cite{our_paper}.

\section{Results}
\label{sec:results}

In order to confirm that both quantities, $\Sigma_Q$ in
Eq.~\eqref{eq:orderparameter_quark} and $\Sigma_S$ in
Eq.~\eqref{eq:orderparameter_scalar}, are well-defined order parameters
we investigate both theories, QCD and  scalar QCD, at finite temperatures.

\begin{figure*}[t!]
  \centering 
  \subfigure{\includegraphics[width=\twofigs]{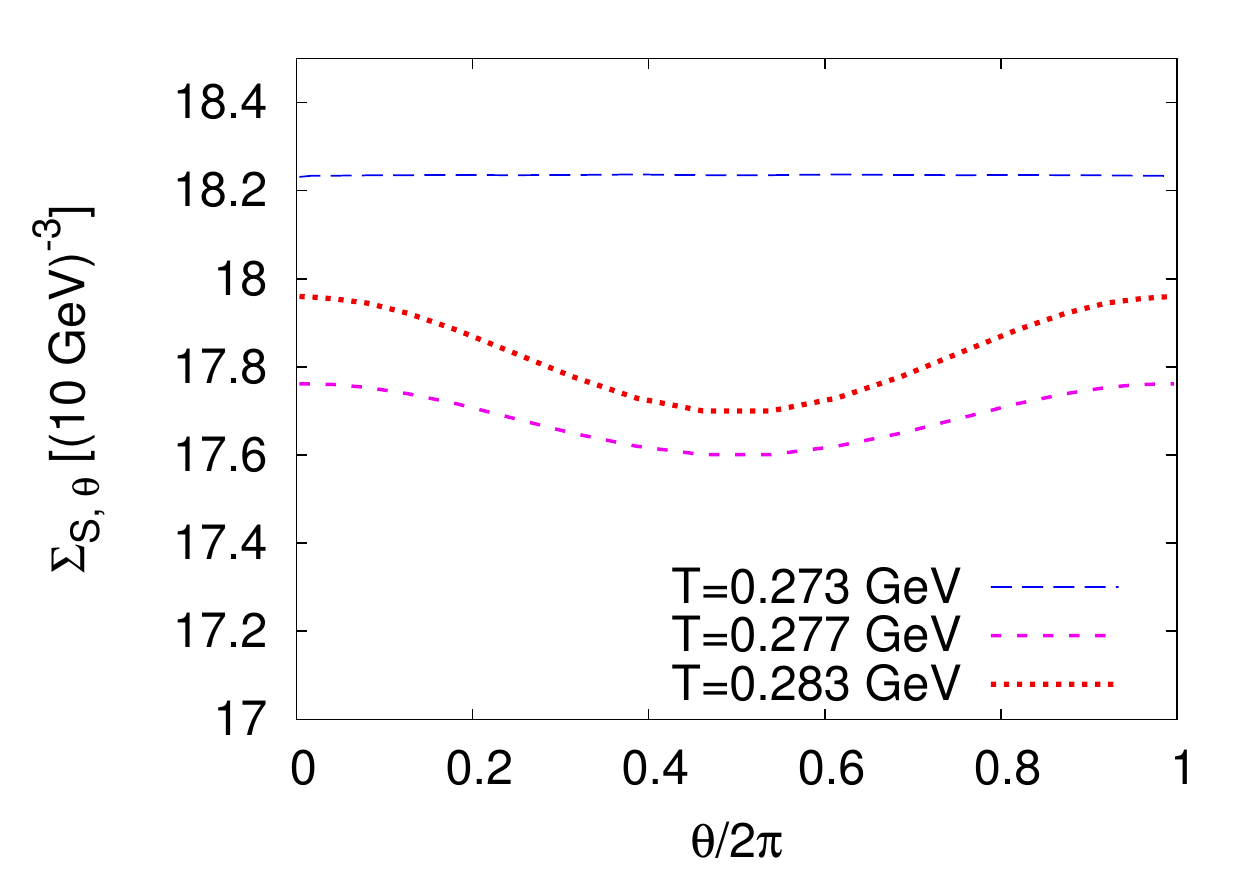}}
  \subfigure{\includegraphics[width=\twofigs]{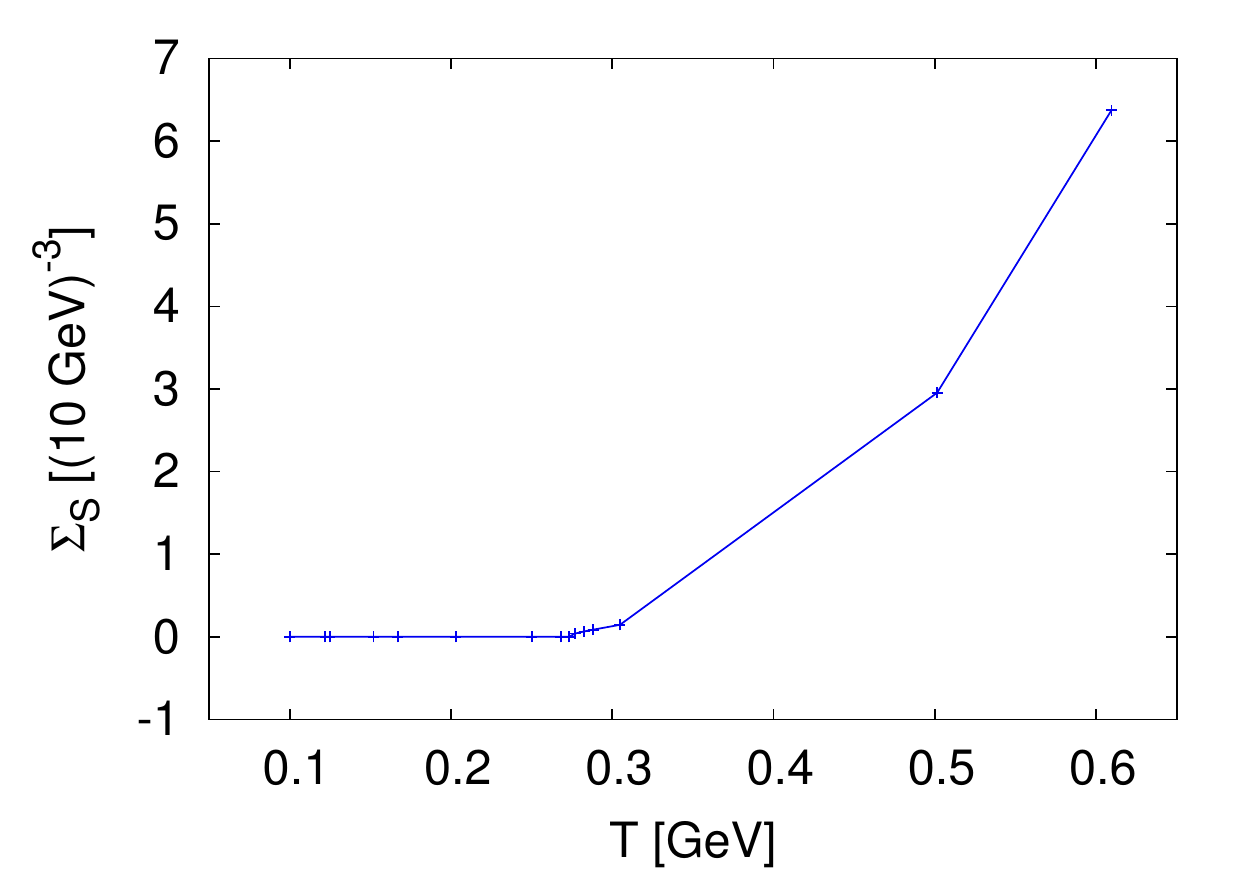}}
  \caption{\label{fig:dse_dual_phi} Left: $\Sigma_{S,\theta}$ of
    Eq.~\protect\eqref{eq:orderparameter_scalar} as a function of the
    boundary conditions for temperatures around the transition. Right: The dual condensate
    $\Sigma_{S}$ as a function of the temperature ($\mu = 4$ GeV and $m = 1.5$ GeV).}
\end{figure*}
\begin{figure*}[t!]
  \centering
  \subfigure{\includegraphics[width=\twofigs]{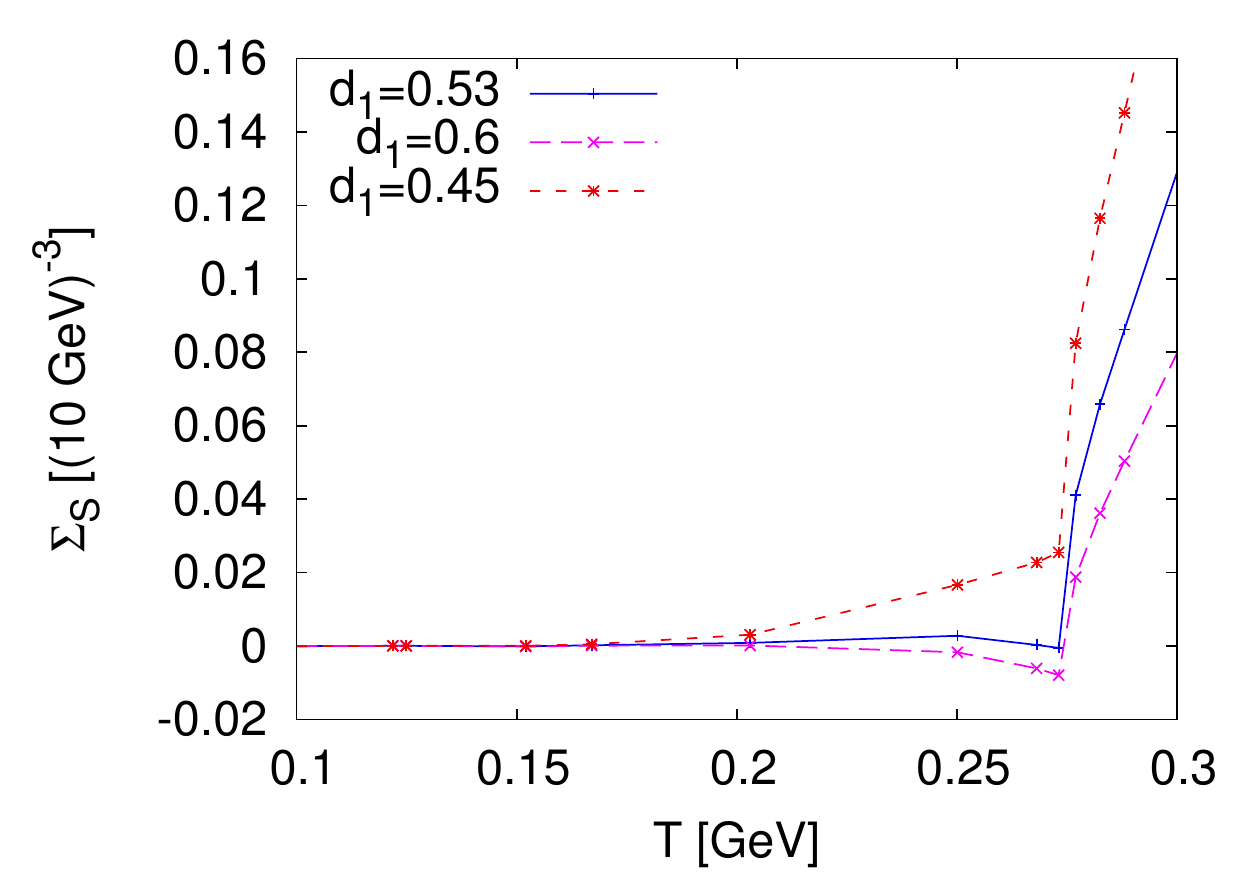}}
  \hspace{-0.4cm}
  \subfigure{\includegraphics[width=\twofigs]{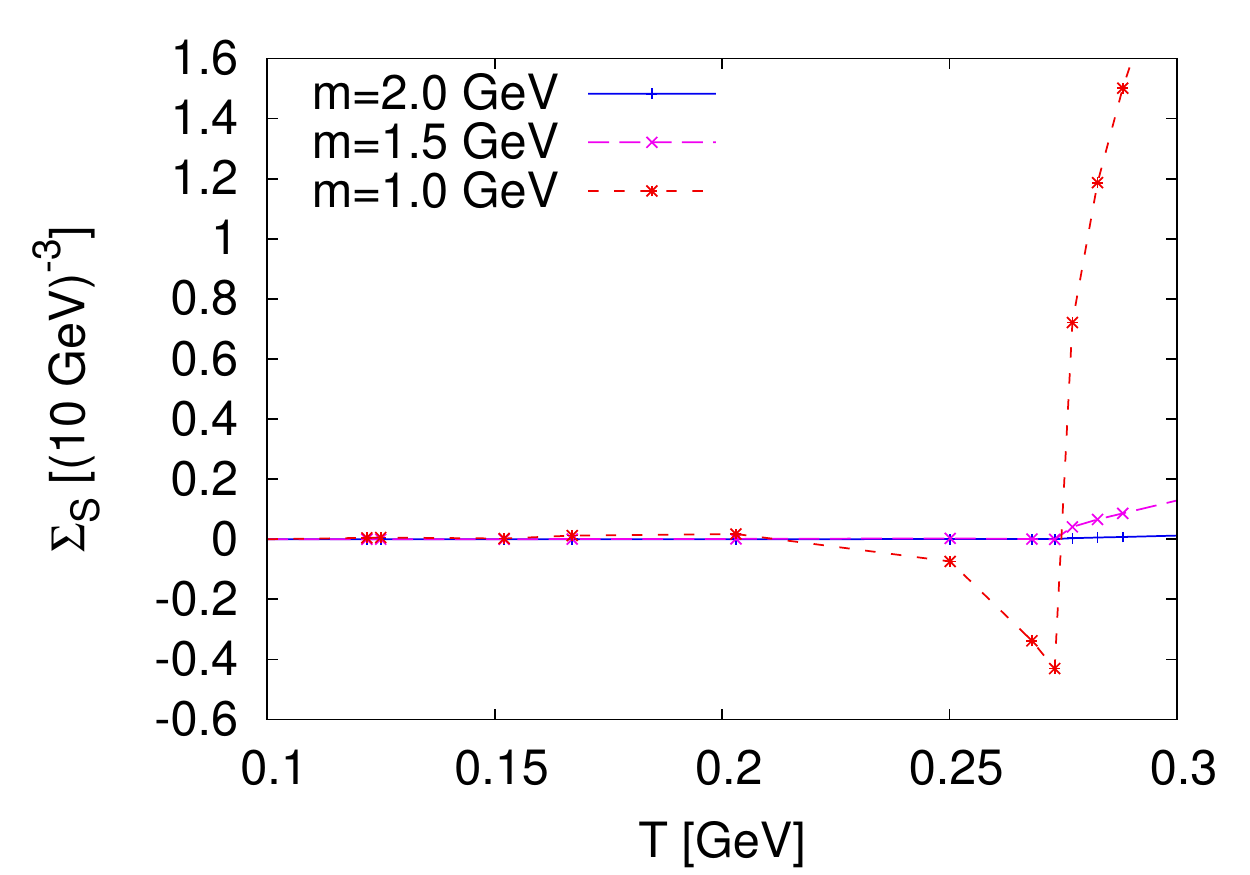}}
  \caption{\label{fig:orderparameter_scalar} Parameter dependency of the dual condensate 
    $\Sigma_S$ as a function of temperature for three different values
    of the vertex parameter $d_1$ (left) and  for three different values of the
    mass of the scalar field (right); ($\mu = 4$ GeV and $m=1.5$ GeV).}
\end{figure*}
In the left panel of \Fig{fig:dse_dual_phi} the $\theta$-dependence of
$\Sigma_{S,\theta}$ for the quenched scalar theory is shown for
temperatures around the transition. From the definition
Eq.~\eqref{eq:orderparameter_scalar} it is clear that $\Sigma_S$
vanishes as long as the $\Sigma_{S,\theta}$ is constant while a
$\theta$-dependency is necessary for a non-vanishing order
parameter. However, these findings for $\Sigma_{S,\theta}$ are
qualitatively similar to QCD with finite quark masses
\cite{Gattringer:2006ci}. In the right panel of \Fig{fig:dse_dual_phi}
the dual condensate $\Sigma_S$ is shown as a function of the
temperature which nicely demonstrates its property as an order
parameter for the center symmetry. Below the transition temperature of
the quenched theory around $T_c\approx 277$ MeV it vanishes and is
finite at higher temperatures. In the vicinity of the critical
temperature the temperature behavior of the order parameter depends
crucially on the model parameters used for the scalar-gluon vertex.
This is demonstrated in \Fig{fig:orderparameter_scalar}, where the
temperature behavior of $\Sigma_S$ is shown for different values of
the $d_1$ parameter in the scalar-gluon vertex
Eq.~\eqref{eq:scalargluonvertex} (left panel) and for various values
of the mass of the scalar field (right panel). Stronger deviations
from a vanishing order parameter slightly below the transition
temperature are observed for smaller mass.  This indicates that the
vertex sensitivity increases towards smaller masses.

\begin{figure*}[t!]
 \centering
 \subfigure{\includegraphics[width=\twofigs]{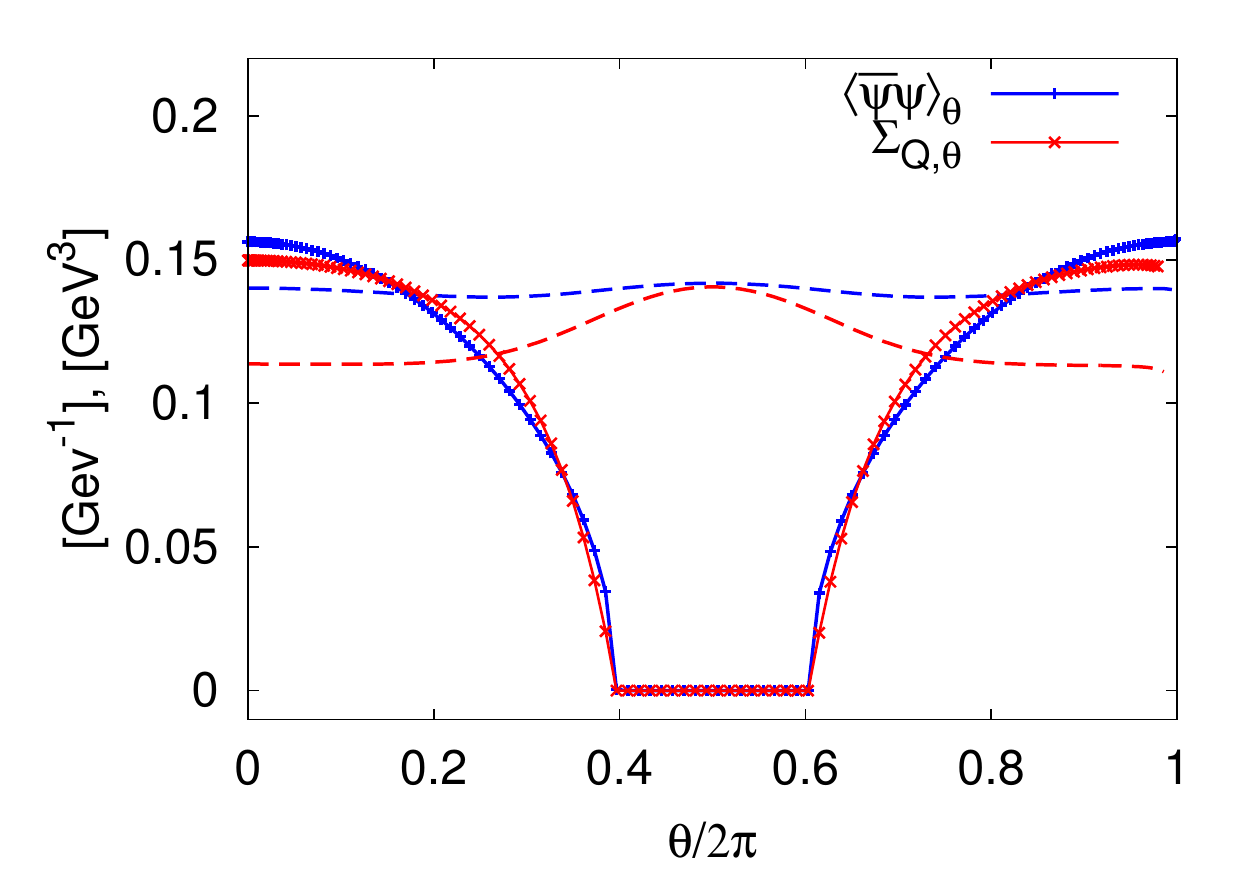}}
 \hspace{-0.4cm}
 \subfigure{\includegraphics[width=\twofigs]{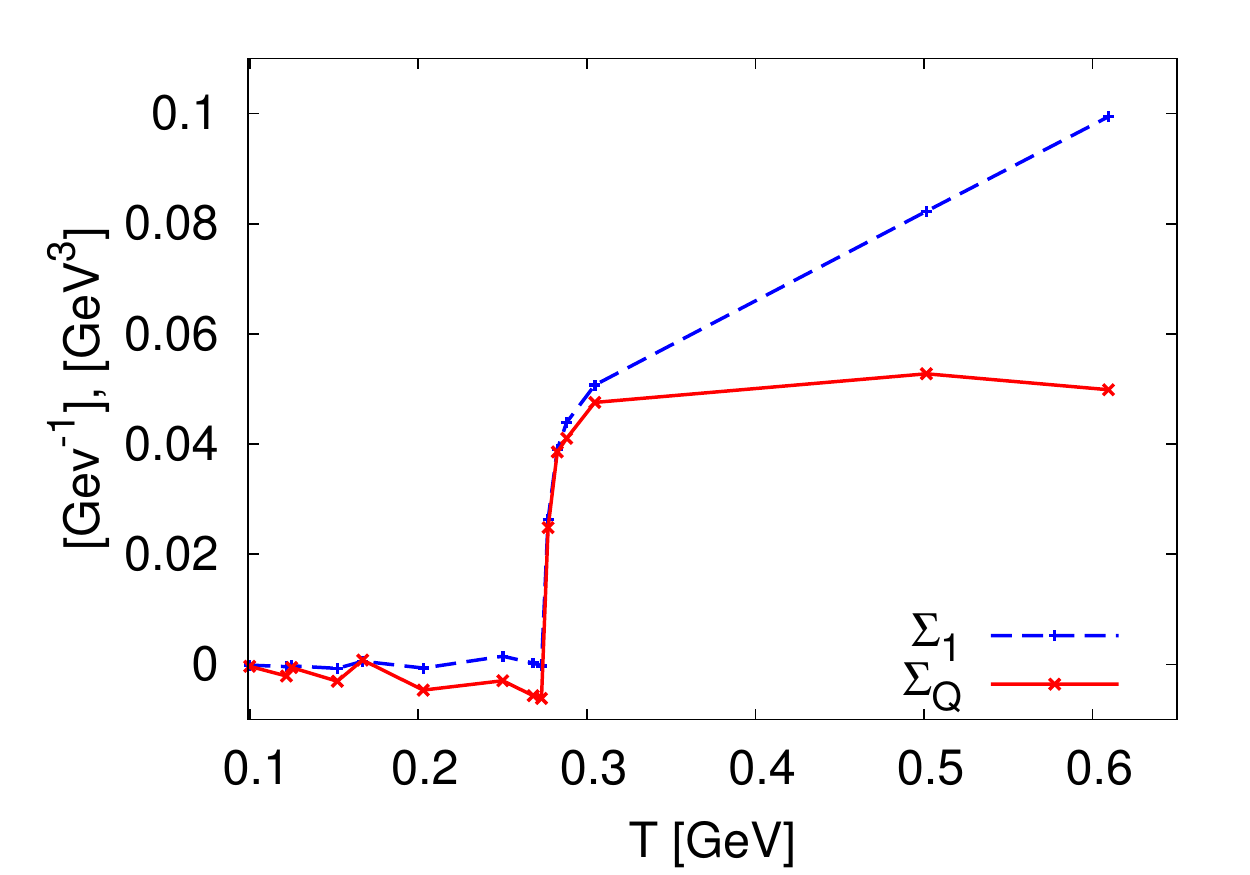}}
 \caption{\label{fig:orderparameter_quark} Left panel: The quark
   condensates $\langle\bar\psi\psi\rangle_\theta$ and
   $\Sigma_{Q,\theta}$, as defined in
   Eq.~\protect\eqref{eq:orderparameter_quark}, as a function of the
   boundary angle for different temperatures in the chiral limit
   (dashed lines: $T=273$ MeV, solid lines: $T=283$ MeV). Right panel:
   The order parameters $\Sigma_{Q}$ and $\Sigma_1$ as defined in
   Eq.~\protect\eqref{eq:dualquarkcondensate} and
   \protect\eqref{eq:orderparameter_quark} as a function of the
   temperature.  }
\end{figure*}

Finally, the proposed new order parameter for ordinary QCD shows a
similar behavior which is presented in the left panel of
\Fig{fig:orderparameter_quark}. In the figure the $\theta$-dependency
of $\Sigma_{Q,\theta}$ in comparison to the chiral condensate is
plotted as a function of the generalized boundary conditions for two
different temperatures.  The results have been obtained in the chiral
limit of the quenched theory. Above the transition temperature,
$\Sigma_{Q,\theta}$ shows the same characteristic plateau as the
chiral condensate for $\theta$'s close to the physical antiperiodic
boundary conditions $\theta=\pi$. Due to the restoration of chiral
symmetry the condensate vanishes which also affects
$\Sigma_{Q,\theta}$ through the scalar dressing function $B(p)$ in the
quark propagator.  The slight variations of $\Sigma_Q$ in the center
symmetric phase below $T_c$ results from a non-constant
$\Sigma_{Q,\theta}$ and can be attributed to lattice artifacts as well
as the choice of the quark-gluon vertex model. However, this effect is
more pronounced in $\Sigma_{Q,\theta}$ than in the chiral condensate.
The right panel of \Fig{fig:orderparameter_quark} shows the
corresponding order parameter $\Sigma_Q$ in comparison to the dual
chiral condensate. Both vanish below $T_c$ and jump immediately to
non-vanishing values above $T_c$. Their deviation at larger
temperatures can be assigned to different dimensionalities.

\section{Conclusions}

We investigated the center phase transition of QCD as well as
fundamentally charged scalar QCD in a quenched formulation. Novel
order parameters are proposed along the lines of previously
constructed dual observables accessible by functional methods.
Solving the Dyson-Schwinger equations for the corresponding matter
propagators numerical results for these order parameters are
presented. A parameter dependency of the employed matter-gluon
vertices on the presented results is found and motivates a more
detailed investigation of its temperature dependence (see also
\cite{Windisch:2012de} for corresponding investigations at
vanishing temperature).

\section*{Acknowledgements}
We are grateful to the organizers of the {\it Xth Quark Confinement and the Hadron
Spectrum} conference for all their efforts which made this extraordinary event
possible.\\
We thank F. Bruckmann, C.S. Fischer, L. Fister, J. Luecker, A. Maas, P. Maris,
J.M. Pawlowski and L. von Smekal for valuable discussions. This work is supported by
the FWF through DK W1203-N16 and grant  P24780-N27.

\end{document}